\documentstyle[11pt,epsfig]{article}
\begin{document}
\title{\bf Dark Energy From Fifth Dimensional Brans--Dicke Theory}
\author{Amir F. Bahrehbakhsh$^{1,2}$\thanks{email: af-bahrehbakhsh@sbu.ac.ir}, Mehrdad Farhoudi$^{1}$\thanks{email:
 m-farhoudi@sbu.ac.ir}\, and Hajar Vakili$^{3}$\thanks{email:
 vakili@physics.sharif.edu} \\
 {\small $^{1}$Department of Physics, Shahid Beheshti University, G.C., Evin, Tehran 19839, Iran} \\
 {\small $^{2}$Department of Physics, Faculty of Science, Payam-e-Nour University, Iran}\\
 {\small $^{3}$Department of Physics, Sharif University of Technology, P.O. Box 11365-9161, Tehran, Iran}}
\date{\small June 4, 2013}
\maketitle
\begin{abstract}
\noindent
 Following the approach of the induced--matter theory, we
investigate the cosmological implications of a five--dimensional
Brans--Dicke theory, and propose to explain the acceleration of
the universe. After inducing in a four--dimensional hypersurface,
we classify the energy--momentum tensor into two parts in a way
that, one part represents all kind of the matter (the baryonic and
dark) and the other one contains every extra terms emerging from
the scale factor of the fifth dimension and the scalar field,
which we consider as the energy--momentum tensor of dark energy.
We also separate the energy--momentum conservation equation into
two conservation equations, one for matter and the other for dark
energy. We perform this procedure for different cases, without
interacting term and with two particular (suitable) interacting
terms between the two parts. By assuming the parameter of the
state equation for dark energy to be constant, the equations of
the model admit the power--law solutions. Though, the
non--interacting case does~not give any accelerated universe, but
the interacting cases give both decelerated and accelerated
universes. For the interacting cases, we figure out analytically
the acceptable ranges of some parameters of the model, and also
investigate the data analysis to test the model parameter values
consistency with the observational data of the distance modulus of
$580$ SNe Ia compiled in Union2.1. For one of these interacting
cases, the best fitted values suggest that the Brans--Dicke
coupling constant $(\omega)$ is $\simeq-7.75$, however, it also
gives the state parameter of dark energy $(w_{_{X}})$ equal to
$\simeq-0.67$. In addition, the model gives the Hubble and
deceleration parameters at the present time to be $H_{\circ}\simeq
69.4$ (km/s)/Mpc and $q_{\circ}\simeq-0.38$ (within their
confidence intervals), where the scale factor of the fifth
dimension shrinks with the time.
\end{abstract}
\bigskip
\newpage
\section{Introduction}

\indent

While improving and searching for more suitable gravitational
theory, attempts for a geometrical unification of gravity with
other interactions has begun by employing extra dimensions beyond
the conventional four--dimensional ({\mbox{$\bf 4D$}})
space--time. At first, Nordstr{\o}m~\cite{Nordstrom} built a
unified theory based on extra dimensions, then
Kaluza~\cite{Kaluza} and Klein~\cite{Klein} established a $5D$
version of general relativity (\textbf{GR}) in which
electrodynamics rises from an extra fifth dimension. Till now, an
intensive amount of researches have been performed on a similar
idea either via different mechanism of compactification of extra
dimension or generalizing it to non--compact
scenarios~\cite{OverduinWesson1997} such as the Brane--World
theories~\cite{Pavsic}, the space--time--matter or induced--matter
(\textbf{IM}) theories~\cite{WessonA,WessonB}, and higher
dimensional cosmology, e.g. Ref.~\cite{Leon}. The latter theories
are grounded on the Campbell--Magaard theorem that asserts every
analytical $N$--dimensional Riemannian space can locally be
embedded in an $(N+1)$--dimensional vacuum
one~\cite{Campbell}--\cite{SeahraWesson}. The importance of this
theorem is that the matter sources of $4D$ space--times can be
viewed as a manifestation of extra dimensions. In another word,
$4D$ field equations with matter sources can be achieved by
inducing $5D$ field equations without matter sources. This idea
has been the core of the IM theory using GR as the underlying
theory.

On the other hand, Jordan~\cite{Jordan} conceived a new kind of
gravitational theory, known as the scalar--tensor theory, by
embedding a curved $4D$ space--time in a flat $5D$ one. Following
his idea, Brans and Dicke~\cite{Brans-Dicke,Dicke} introduced a
version of the scalar--tensor gravitational theory, as an
alternative to GR, with a non--minimally scalar field coupled to
the curvature, in which the weak equivalence principle is
preserved and makes it to be more Machian than GR. Unfortunately,
its weakness is mismatching with the solar system
observations~\cite{Bertotti2003,fujiBook}, which is a generic
difficulty of the scalar--tensor theories in the solar system
constraint~\cite{banerjee2001,Sen2001}. However, it does~not
necessarily denote that the evolution of the universe, at all
scales, should be close to GR, in which there are some debates on
its tests on the cosmic scales as well~\cite{bean2009,Danieletal}.

The measurements of anisotropies in the cosmic microwave
background suggest that the ordinary $4D$ universe is very close
to a spatially flat universe~\cite{Bachcall}--\cite{komatsu2011},
and the observations of type Ia--supernovae indicate that the
expansion of the universe presently is
accelerating~\cite{Perlmutter}--\cite{Carroll}. Hence, in one
school of thought, the main component of the universe should be
consist of what usually has been called dark
energy~\cite{DarkEnergyA,DarkEnergyB}. Since then, a considerable
amount of work has been performed in the literature to explain the
acceleration of the universe. Most of the dark energy models, such
as the quintessence~\cite{QuintessenceA}--\cite{QuintessenceD},
the chaplygin gas~\cite{ChaplyginA,ChaplyginB} and the
k-essence~\cite{K-essenceA,K-essenceB} models, involve minimally
coupled scalar fields with different potentials. These scalar
fields or the potentials have been added in \emph{priori} by hand,
and hence, their origins are~not understood. Nevertheless, in the
recent decades, explaining the accelerated expansion of the
universe via fundamental theories has been a great challenge.

Recently, a $5D$ vacuum Brans--Dicke (\textbf{BD}) theory based on
the idea of IM theory has been investigated~\cite{Aguilar}, in
which the role of GR has been replaced by the BD theory of
gravitation as the fundamental underlying theory. It has been
illustrated that $5D$ vacuum BD equations, when reduced to four
dimensions, give a modified version of the $4D$ BD theory with an
induced potential. Whereas in the literature, to obtain
accelerating universes, such potentials has been added in
\emph{priori} by hand. A few applications of this approach have
also been performed in Refs.~\cite{Ponce1}--\cite{Rasouli}.

Although one of the aims of higher dimensional theories is to
obtain $4D$ matter from pure geometry, but it is sometimes
desirable to have a higher dimensional energy--momentum tensor or
a scalar field, for example in compactification of the extra
curved dimensions~\cite{Appelquist}. Indeed, the reduction
procedure of a $5D$ analogue of the BD theory with matter content,
on every hypersurface orthogonal to an extra cyclic dimension
(that recovers the modified BD theory described by a 4--metric
coupled to two scalar fields) has previously been accomplished in
the literature~\cite{qiang2005,qiang2009}. Besides, the main idea
of the brane models is the existence of a higher dimensional bulk
in which the universe is sitting as a hypersurface and all the
matter fields, except gravity, are confined to this
brane~\cite{Pavsic}. Cosmological implications in the context of
brane worlds have also been studied, e.g. it has been shown that
the recent accelerated expansion of the universe can be explained
by a geometrical originated dark energy caused by an addition of a
brane curvature scalar term in the action~\cite{BraneA,BraneB}.

Similar to the idea of brane scenarios, but based on the IM theory
with underlying the BD theory of gravitation, in this work, we
employ a generalized Friedmann--Lema\^{\i}tre--Robertson--Walker
(\textbf{FLRW}) type solution for a $5D$ BD theory and investigate
its cosmological implications. For this purpose, in the next
section, we give a brief review of a $5D$ BD theory following the
idea of IM theory. Then, in a $4D$ hypersurface, we gather the
extra terms emerging from the scale factor of the fifth dimension
and the scalar field as dark energy component of an
energy--momentum tensor in addition to the usual energy--momentum
tensor of all (the baryonic and dark) matter. In Section~$3$, we
consider a generalized FLRW metric in a $5D$ space--time, and
derive the FLRW cosmological equations. Then, we manage and obtain
the total energy conservation equation as separated into two
energy conservation equations for all matter and dark energy, with
non--interacting and two simple interacting cases. However, as the
cosmological equations are strictly non--linear and do~not have
exact solution, in order to proceed, we assume some
simplifications. In Section~$4$, we apply the observational
constraints for the interacting cases of the model, and use the
Markov Chain Monte Carlo (\textbf{MCMC}) method to fit the free
parameters of the model with the SCP Union2.1 SN Ia
compilation~\cite{suzuki2011,supernova} based on the Bayesian
statistics~\cite{AmirHajian}. A few tables and figures are also
provided for a better view of the acceptable range and the best
fitted values of parameters. Finally, conclusions are presented in
the last section.

\section{Five--Dimensional Brans--Dicke Theory}
\indent

Following the idea of IM theories, one can consider the BD theory
of gravitation as the underlying theory instead of GR. In this
respect, the action of $5D$ BD theory in the Jordan frame can
analogously be written as
\begin{equation}\label{5D Action}
\emph{S} \\\
[g_{_{AB}},\phi]=\int\sqrt{|{}^{_{(5)}}g|} \left (\phi \
^{^{(5)}}\!R-\frac{\omega}{\phi}g^{_{AB}}\phi_{,_{A}}\phi_{,_{B}}+
16\pi L_{m} \right )d^{5}x\, ,
\end{equation}
where $c=1$, the capital Latin indices run from zero to four,
$^{^{(5)}}R$ is $5D$ Ricci scalar, $^{_{(5)}}g$ is the determinant
of $5D$ metric $g_{_{AB}}$, $\phi$ is a positive scalar field that
describes the gravitational coupling in five dimensions,
$\emph{L}_{m}$ represents the matter Lagrangian and $\omega$ is a
dimensionless coupling constant. Hence, the field equations
obtained from action (\ref{5D Action}) are
\begin{equation}\label{5D Einstein Eq}
^{^{(5)}}G_{_{AB}}=\frac{8\pi}{\phi} \
^{^{(5)}}T_{_{AB}}+\frac{\omega}{\phi^2}
\left(\phi_{,_{A}}\phi_{,_{B}}-\frac{1}{2}g_{_{AB}}\phi^{,_{C}}\phi_{,_{C}}
\right)+\frac{1}{\phi}\left(\phi_{;_{AB}}-g_{_{AB}}\
^{^{(5)}}\Box\phi \right)
\end{equation}
and
\begin{eqnarray}\label{5D Scalar Eq}
^{^{(5)}}\Box\phi=\frac{8\pi}{4+3\omega}\ ^{^{(5)}}T\, ,
\end{eqnarray}
where $^{^{(5)}}G_{_{AB}}$ is $5D$ Einstein tensor,
$^{^{(5)}}T_{_{AB}}$ is $5D$ energy--momentum tensor, $^{^{(5)}}T
\equiv\ ^{^{(5)}}T^{^{C}}{}_{^{C}}$ and $^{^{(5)}}\Box\!
\equiv{}_{;_{A}}{}^{^{A}}$. In order to have a non--ghost scalar
field in the conformally related Einstein frame, i.e. a field with
a positive kinetic energy term in that frame, the BD coupling
constant must be $\omega>-4/3$~\cite{Freund}--\cite{Barrowetal}.
Though recently, some new ghost dark energy models have been
suggested to explain the observed acceleration of the
universe~\cite{Urban}.

As a plausible assumption, we consider that $^{^{(5)}}T_{_{AB}}$
exactly represents the same baryonic and dark matter sources of a
$4D$ hypersurface, i.e. $T^{^{(M)}}_{\alpha\beta}$. Hence, in this
case $^{^{(5)}}T_{_{AB}}= diag
(\rho_{_{M}},-\emph{p}_{_{M}},-\emph{p}_{_{M}},-\emph{p}_{_{M}},0)$,
where $\rho_{_{M}}$ and $p_{_{M}}$ are the energy density and the
pressure of the matter, and the Greek indices run from zero to
three. Also, cosmological purposes usually restrict attentions to
$5D$ metrics of the simple form
\begin{equation}\label{5D Metric}
dS^2=g_{_{AB}}(x^{C})dx^{A}dx^{B}=\
^{^{(5)}}\!g_{\mu\nu}(x^{C})dx^{\mu}dx^{\nu}+g_{_{44}}(x^{C})dy^2
\equiv \ ^{^{(5)}}\!g_{\mu\nu}(x^{C})dx^{\mu}dx^{\nu}+\epsilon
b^{2}(x^{C})dy^2
\end{equation}
in local coordinates $x^{A}=(x^{\mu},y)$,
where $y$ represents the fifth coordinate and $\epsilon^2=1$.
By assuming the $5D$ space--time is foliated
by a family of hypersurfaces, say $\Sigma$, that are defined by fixed values of
$y$, then, one can obtain the metric intrinsic to every generic
hypersurface, e.g. $\Sigma_{\circ}(y=y_{\circ})$, by
restricting the line element (\ref{5D Metric}) to displacements confined
to it. Thus, the induced metric on the hypersurface $\Sigma_{\circ}$
becomes
\begin{equation}\label{4D Metric}
ds^2=\
^{^{(5)}}\!g_{\mu\nu}(x^{\alpha},y_{\circ})dx^{\mu}dx^{\nu}\equiv
g_{\mu\nu}dx^{\mu}dx^{\nu}\, ,
\end{equation}
in such a way that the usual $4D$ space--time metric,
$g_{\mu\nu}$, can be recovered.

Therefore, after some manipulations, equation (\ref{5D Einstein
Eq}) on the hypersurface $\Sigma_{\circ}$ can be written as
\begin{equation}\label{4D Einstein Eq}
G_{\alpha\beta}=\frac{8\pi}{\phi}(T^{^{(M)}}_{\alpha\beta}+T^{^{(X)}}_{\alpha\beta}),
\end{equation}
where we consider $T^{^{(X)}}_{\alpha\beta}$ as dark energy
component of the energy--momentum tensor that is defined by
\begin{equation}\label{T Dark Energy}
T^{^{(X)}}_{\alpha\beta}\equiv T^{^{\rm(IM)}}_{\alpha\beta}+T^{^{(\phi)}}_{\alpha\beta}+\frac{\omega}{8\pi\phi}
 \Big (\phi_{,\alpha}\phi_{,\beta}-\frac{1}{2}g_{\alpha\beta}\phi^{,\sigma}\phi_{,\sigma} \Big )
+\frac{1}{8\pi}\big (\phi_{;\alpha\beta}-g_{\alpha\beta} \Box\phi\big ),
\end{equation}
in which, analogous to the IM theory~\cite{Amirfarshad},
\begin{eqnarray}\label{T IM}
T^{^{\rm(IM)}}_{\alpha\beta}\equiv\frac{\phi}{8\pi}
\Bigg \{  \frac{b_{;\alpha\beta}}{b}-\frac{\Box b}{b}g_{\alpha\beta}-\frac{\epsilon}{2b^2}
\Bigg [\frac{b'}{b}g'_{\alpha\beta} -g''_{\alpha\beta}+g^{\mu\nu}g'_{\alpha\mu}
g'_{\beta\nu}-\frac{1}{2}g^{\mu\nu}g'_{\mu\nu}g'_{\alpha\beta}
\nonumber\\
-g_{\alpha\beta}\bigg (\frac{b'}{b}g^{\mu\nu}g'_{\mu\nu}-g^{\mu\nu}g''_{\mu\nu}-\frac{1}
{4}g^{\mu\nu}g^{\rho\sigma}g'_{\mu\nu}g'_{\rho\sigma}
-\frac{3}{4}g'^{\mu\nu}g'_{\mu\nu} \bigg )\Bigg ]
\Bigg \}
\end{eqnarray}
and
\begin{equation}\label{T Phi}
T^{^{(\phi)}}_{\alpha\beta}\equiv-\frac{\epsilon}{8\pi b^2}\Bigg
\{g_{\alpha\beta}\Bigg [\phi''+\Big (\frac{1}{2}g^{\mu\nu}g'_{\mu\nu}-\frac{b'}{b}+\frac{\omega}{2}\frac{\phi'}{\phi}\Big )
\phi'+\epsilon bb_{,\mu}\phi^{,\mu} \Bigg ]-\frac{1}{2}g'_{\alpha\beta}\phi'  \Bigg \}.
\end{equation}
The prime denotes derivative with respect to the fifth coordinate.

In the following section, we consider a generalized FLRW metric in
a $5D$ universe and investigate its cosmological properties.

\section{Generalized FLRW Cosmology}
\indent

For a $5D$ universe with an extra space--like dimension in
addition to the three usual spatially homogenous and isotropic
ones, metric (\ref{5D Metric}), as a generalized FLRW solution,
can be written as
\begin{equation}\label{FRW Metric}
dS^2=-dt^2+a^2(t)\left [\frac{dr^2}{1-kr^2}+r^2(d\theta^2+\sin^2\theta
d\varphi^2)\right ]+b^2(t)dy^2\,.
\end{equation}
Generally, the scalar field $\phi$ and the scale factors $a$ and
$b$ should be functions of $t$ and $y$. However, for physical
plausibility and simplicity, we assume that the
hypersurface--orthogonal space--like is a Killing vector field in
the underlying $5D$ space--time~\cite{qiang2005,qiang2009}, i.e.
the extra dimension to be a cyclic coordinate. Besides, the
functionality of the scale factor $b$ on $y$ could be eliminated
either by transforming to a new extra coordinate (if $b$ be a
separable function) or making no changes in the consequent
equations if $b$ is the only field that depends on $y$. In another
word, in the compactified extra dimension scenarios, all fields
can be Fourier--expanded around the fixed value $y_{\circ}$, and
hence, one can get the observable terms independent of $y$, i.e.
physics would be effectively independent of the compactified fifth
dimension, see, e.g. Ref.~\cite{OverduinWesson1997}. With this
solution, we will show that the universe can accept both
accelerating and decelerating expansion eras.

By considering metric (\ref{FRW Metric}), the BD equations
(\ref{5D Einstein Eq}) reduce as follows. The time component
$A=0=B$ gives
\begin{equation}\label{FRW1}
H^2=\frac{8\pi}{3\phi}\rho_{_{M}}+\frac{\omega}{6}F^2-HB+\frac{\ddot{\phi}}{3\phi}-\frac{k}{a^2}\equiv
\frac{8\pi}{3\phi}(\rho_{_{M}}+\rho_{_{X}})-\frac{k}{a^2}\, ,
\end{equation}
the spatial components $A=B=1,2,3$ provide
\begin{equation}\label{FRW2}
\frac{\ddot{a}}{a}=-\frac{4\pi}{\phi}p_{_{M}}-\frac{1}{2}H^{2}-\frac{\omega}{4}F^{2}+\frac{1}{2}HF
-HB-\frac{\ddot{b}}{2b}-\frac{k}{2a^2}\equiv
-\frac{4\pi}{3\phi} \Big[(\rho_{_{M}}+\rho_{_{X}})+3(p_{_{M}}+p_{_{X}}) \Big]\,
\end{equation}
and the $A=4=B$ component yields
\begin{equation}\label{FRW3}
\frac{\ddot{a}}{a}=-H^{2}-\frac{\omega}{6}F^{2}+\frac{1}{3}BF-\frac{k}{a^2}\,.
\end{equation}
Also, the scalar field equation (\ref{5D Scalar Eq}) becomes
\begin{equation}\label{4D Scalar Eq}
\frac{\ddot{\phi}}{\phi}+3HF=\frac{8\pi(\rho_{_{M}}-3p_{_{M}})}{\phi(4+3\omega)}-BF\, ,
\end{equation}
where $H\equiv\dot{a}/a$, $B\equiv\dot{b}/b$ and $F\equiv\dot{\phi}/\phi$.

In the last part of equations (\ref{FRW1}) and (\ref{FRW2}), we
have defined the energy density and pressure of dark energy as
\begin{equation}\label{Ro X}
\rho_{_{X}}\equiv -T^{^{(X)}t}{}_{t}=\frac{\phi}{8\pi}\Big
(\frac{\omega}{2}F^2-3HB+\frac{\ddot{\phi}}{\phi} \Big )
\end{equation}
and
\begin{equation}\label{P X}
p_{_{X}}\equiv
T^{^{(X)}i}{}_{i}=\frac{\phi}{8\pi}\Big(\frac{\omega}{2}F^{2}-HF+2HB+\frac{\ddot{b}}{b}
\Big)\, .
\end{equation}
Hence, the state equation of dark energy obviously yields
\begin{equation}\label{W X}
w_{_{X}}=\frac{p_{_{\rm X}}}{\rho_{_{\rm X}}}=\frac{\omega F^{2}/2-HF+2HB+\ddot{b}/b}{\omega F^{2}/2-3HB+\ddot{\phi}/\phi}\, .
\end{equation}

In order to find the functionality of $w_{_{X}}$ with the time,
equations (\ref{FRW1})--(\ref{4D Scalar Eq}) must be exactly
solved. However, as these equations are coupled non--linearly, one
should apply numerical methods. As an alternative, in the
following, by considering a few simplifications, we proceed to fit
the parameters of the model by observational data analysis, and
discuss the properties of the model.

First of all, let us obtain the energy conservation equations. In
this respect, one can take the time derivative of equation
(\ref{FRW1}) and substitutes equation (\ref{FRW2}) into it, and
finally gets
\begin{equation}\label{Conservation MIX}
\dot{\tilde{\rho}}+3H(\tilde{\rho}+\tilde{p})=F\tilde{\rho}\, ,
\end{equation}
where $\tilde{\rho}\equiv\rho_{_{M}}+\rho_{_{X}}$ and
$\tilde{p}\equiv p_{_{M}}+p_{_{X}}$. As the nature of dark energy
is unknown, the detailed coupling form among it and matter is
unclear. Hence, one can expect that their conservation equations
should not be independent. In this case, let us apply a plausible
simplification by separating equation (\ref{Conservation MIX})
into two distinguished continuity equations for $\rho_{_{X}}$ and
$\rho_{_{M}}$ as
\begin{equation}\label{Conservation M}
\dot{\rho}_{_{M}}+3H(\rho_{_{M}}+p_{_{M}})=F\rho_{_{M}}+Q
\end{equation}
and
\begin{equation}\label{Conservation X}
\dot{\rho}_{_{X}}+3H(\rho_{_{X}}+p_{_{X}})=F\rho_{_{X}}-Q\, ,
\end{equation}
where the $Q$ term stands for interacting terms among matter and
dark energy. Such a similar interacting term has also been used in
the literature, e.g. Ref.~\cite{Movahed}.

To proceed further, we take $w_{_{M}}=0$ as dust (the baryonic and
dark) matter, and assume $w_{_{X}}$ to be a constant. Hence,
relation (\ref{W X}) imposes the power--law solutions
\begin{equation}\label{a}
a(t)=a_{\circ}\left(\frac{t}{t_{\circ}}\right)^{\alpha} \qquad {\rm with} \qquad  H=\frac{\alpha}{t}\, ,
\end{equation}
\begin{equation}\label{b}
b(t)=b_{\circ}\left(\frac{t}{t_{\circ}}\right)^{\beta} \qquad {\rm with} \qquad B=\frac{\beta}{t}
\end{equation}
and
\begin{equation}\label{c}
\phi(t)=\phi_{\circ}\left(\frac{t}{t_{\circ}}\right)^{\gamma} \qquad {\rm with} \qquad F=\frac{\gamma}{t}\, .
\end{equation}

If one assumes the power--law solutions (\ref{a})--(\ref{c}),
equations (\ref{FRW1})--(\ref{FRW3}) will restrict either the
geometry to be spatially flat or $\alpha$ to be one, i.e. a free
expanding universe. As the latter choice is~not interested, we
consider $k=0$ in this work. This choice is also consistent with
the measurements of anisotropies in the cosmic microwave
background radiation that indicate the universe must be very close
to spatially flat one~\cite{Bachcall}--\cite{komatsu2011}.

Note that, there are four independent equations
(\ref{FRW1})--(\ref{4D Scalar Eq}) that, in general, determine
$a$, $b$, $\phi$ (or equivalently $\alpha$, $\beta$, $\gamma$ in
the power--law solutions) and $\rho_{_{M}}$. These unknowns also
depend on the BD parameter $\omega$. In order that our {\it
ans\"{a}tze} do~not lead to over--constraining equations, we
manage the values of the parameters of the model including
$\omega$ to be, by using the data analysis method, consistent with
the observations.

In the next two subsections, we find the relations of
$\rho_{_{X}}$ and $\rho_{_{M}}$ with the scale factors and the
scalar field for both the non--interacting and interacting cases.
For interacting cases, we consider two simple and reasonable
choices of $Q\equiv\Gamma\, \rho_{_{M}}$ and $Q\equiv\Gamma\,
\rho_{_{X}}$, where
\begin{equation}\label{Gamma}
\Gamma\equiv\frac{\dot{\psi}}{\psi}.
\end{equation}
The symbol $\psi$ is denoted as a general notation that represents
either the scale factor of the ordinary spatial dimensions, $a$,
or the scale factor of the fifth dimension, $b$, or the scalar
field, $\phi$. That is, $\Gamma$ can be either $H$ or $B$ or $F$,
however, by matching the model with the observations, we will
figure out the best choice of $\psi$.

\subsection{Non--Interacting Case $Q=0$}
\indent

Now, for the matter to be a dust one and the dark energy parameter
of state to be time independent, if there is no interacting term,
then equations (\ref{Conservation M}) and (\ref{Conservation X})
lead to
\begin{equation}\label{Ro M1}
\frac{\rho_{_{M}}}{\rho_{_{M\circ}}}=\frac{\phi}{\phi_{_{\circ}}} \left (\frac{a}{a_{_{\circ}}} \right )^{-3}
\end{equation}
and
\begin{equation}\label{Ro X1}
\frac{\rho_{_{X}}}{\rho_{_{X\circ}}}=\frac{\phi}{\phi_{_{\circ}}} \left (\frac{a}{a_{_{\circ}}} \right )^{-3(1+w_{_{X}})},
\end{equation}
which for the power--law solutions (\ref{a})--(\ref{c}) become
\begin{equation}\label{Ro M11}
\frac{\rho_{_{M}}}{\rho_{_{M\circ}}}=\left (\frac{t}{t_{_{\circ}}} \right )^{\gamma-3\alpha}
\end{equation}
and
\begin{equation}\label{Ro X11}
\frac{\rho_{_{X}}}{\rho_{_{X\circ}}}=\left (\frac{t}{t_{_{\circ}}} \right )^{\gamma-3\alpha(1+w_{_{X}})}.
\end{equation}
where $\rho_{_{M\circ}}$, $\rho_{_{X\circ}}$, $a_{\circ}$ and
$\phi_{\circ}$ are the energy density of matter, the energy
density of dark energy, the scale factor and the scalar field at
the present time, respectively.

Substituting the power--law solutions in equation (\ref{FRW1})
(and also equation (\ref{4D Scalar Eq})) imposes $\alpha=2/3$ and
$w_{_{X}}=0$, which does~not give an accelerated universe. Hence,
in the following, we investigate different interacting cases to
figure out any possible accelerating solutions. However, when
$w_{_{X}}$ is~not constant (as a more general case), then the
non--interacting choice may also provide some interesting results.
\subsection{Interacting Cases $Q\equiv \Gamma\rho_{_{M}}$ and $Q\equiv \Gamma\rho_{_{X}}$}
\indent

With the same assumptions as in the non--interacting cases, to
find out analytically the ranges of the parameters of our model
for interacting cases $Q\equiv \Gamma\rho_{_{M}}$ and $Q\equiv
\Gamma\rho_{_{X}}$, we consider the power--law solutions
(\ref{a})--(\ref{c}). Thus, we assume
\begin{equation}\label{Si}
\psi(t)=\psi_{\circ}\left(\frac{t}{t_{\circ}}\right)^{\lambda} \qquad {\rm with} \qquad \Gamma=\frac{\lambda}{t}\, .
\end{equation}
Now, for the chosen interacting cases, we will show, in the
following two parts, that the model can accept both acceleration
and deceleration for the universe expansion eras.
\\
\\
\\
\textbf{A. Interacting Case $Q\equiv \Gamma\rho_{_{M}}$}
\\

Substituting $Q\equiv\Gamma\rho_{_{M}}$ into equation (\ref{Conservation M}) gives
\begin{equation}\label{Ro M2}
\frac{\rho_{_{M}}}{\rho_{_{M\circ}}}=\frac{\phi}
{\phi_{_{\circ}}}\frac{\psi}{\psi_{_{\circ}}} \left
(\frac{a}{a_{_{\circ}}} \right )^{-3},
\end{equation}
that for the power--law solutions (\ref{a})--(\ref{c}), becomes
\begin{equation}\label{Ro M3}
\frac{\rho_{_{M}}}{\rho_{_{M\circ}}}=\left (\frac{t}{t_{_{\circ}}} \right )^{\gamma+\lambda-3\alpha}.
\end{equation}
Thus, an acceptable solution of (\ref{Conservation X}), which
satisfies equations (\ref{FRW1})--(\ref{4D Scalar Eq}), is
\begin{equation}\label{Ro X3}
\frac{\rho_{_{X}}}{\rho_{_{X\circ}}}=\left (\frac{t}{t_{_{\circ}}}
\right )^{\gamma+\lambda-3\alpha} \qquad {\rm with} \qquad
\rho_{_{X\circ}}=-\frac{\lambda \rho_{_{M\circ}}}{\lambda+3\alpha
w_{_{X}}}\, .
\end{equation}

Solutions (\ref{Ro M3}) and (\ref{Ro X3}) reveals that both the
matter and dark energy evolutions are the same in this model.
Substituting $\rho_{_{M}}$ and $\rho_{_{X}}$ into equation
(\ref{FRW1}) or (\ref{FRW2}), imposes
\begin{equation}\label{Lambda1}
\lambda=3\alpha-2\, ,
\end{equation}
and then, using it into the second part of (\ref{Ro X3}), gives
\begin{equation}\label{W X1}
w_{_{X}}=-\frac{(u_{\circ}+1) (3\alpha-2) }{3\alpha }\, ,
\end{equation}
where $u_{\circ}\equiv\rho_{_{M\circ}}/\rho_{_{X\circ}}$.
The acceptable ranges of $w_{_{X}}$, $\alpha$ and $\lambda$,
for an expanding universe, are given in Table~$1$ and are depicted in Fig.~$1$.

For this interacting case, equation (\ref{FRW1}) can also be presented as
\begin{equation}\label{HZ2}
H^2=H_{\circ}^2 \Big[\Omega_{M\circ}+\Omega_{X\circ} \Big](1+z)^{\frac{2}{\alpha}},
\end{equation}
where $\Omega_{M\circ}+\Omega_{X\circ}=1$. Relation (\ref{HZ2}) is
the same as the result obtained in Ref.~\cite{Sen} for a $4D$ BD
model with a \emph{priori} assumed (i.e. added--by--hand)
potential term. However, in the next section, we also match the
model consistency with the observational data of SN Ia, and search
the best fitted parameters of the model. If one sets $\psi\equiv
a$, i.e. $\Gamma\equiv H$ and $\lambda\equiv\alpha$, then from
equation (\ref{Lambda1}) it is obvious that $\alpha=1$, which
gives a free expanding universe, thus, in the data analyzing of
equation (\ref{HZ2}), we only consider the other two (more
interesting) cases $\lambda=\beta$ and $\lambda=\gamma$.
\\
\\
\\
\noindent \textbf{B. Interacting Case $Q\equiv \Gamma\rho_{_{X}}$}
\\

Another suitable choice for non--interacting case is to take $Q\equiv\Gamma\rho_{_{X}}$.
Thus, by substituting it into equation (\ref{Conservation X}), one gets
\begin{equation}\label{Ro X4}
\frac{\rho_{_{X}}}{\rho_{_{X\circ}}}=\frac{\phi}{\phi_{_{\circ}}}
\left (\frac{\psi}{\psi_{_{\circ}}} \right )^{-1}  \left
(\frac{a}{a_{_{\circ}}} \right )^{-3(1+w_{_{X}})},
\end{equation}
that for the power--law solutions (\ref{a})--(\ref{c}), yields
\begin{equation}\label{Ro X5}
\frac{\rho_{_{X}}}{\rho_{_{X\circ}}}=\left (\frac{t}{t_{_{\circ}}} \right )^{\gamma-\lambda-3\alpha(1+w_{_{X}})}.
\end{equation}
Therefore, an acceptable solution of (\ref{Conservation M}) is
\begin{equation}\label{Ro M5}
\frac{\rho_{_{M}}}{\rho_{_{M\circ}}}=\left (\frac{t}{t_{_{\circ}}}
\right )^{\gamma-\lambda-3\alpha(1+w_{_{X}})} \qquad {\rm with}
\qquad  \rho_{_{M\circ}}=-\frac{\lambda
\rho_{_{X\circ}}}{\lambda+3\alpha w_{_{X}}}.
\end{equation}

Once again, the matter and dark energy evolve in the same manner
in this model. Solutions (\ref{Ro X5}) and (\ref{Ro M5}) must also
satisfy equations (\ref{FRW1})--(\ref{4D Scalar Eq}), that give
\begin{equation}\label{Lambda2}
\lambda=u_{\circ}(3\alpha-2)
\end{equation}
and
\begin{equation}\label{W X2}
w_{_{X}}=-\frac{(u_{\circ}+1) (3\alpha-2) }{3\alpha }\, .
\end{equation}
It is notable that equations (\ref{W X2}) and (\ref{W X1}) are exactly the same.
Finally, one can rewrite equation (\ref{FRW1}) as
\begin{equation}\label{HZ3}
H^2=H_{\circ}^2 \Big [\Omega_{M\circ}+\Omega_{X\circ} \Big](1+z)^{3(1-u_{\circ})+\frac{2u_{\circ}}{\alpha}},
\end{equation}
where again $\Omega_{M\circ}+\Omega_{X\circ}=1$. In the following,
we highlight the conditions that correspond to a decelerating and
an accelerating universe for both of the interacting cases.
\begin{description}
\item \textbf{I. Deceleration and Free Expansion}\\

The observational data reveals that when the universe was in the
radiation or dust dominated phases, it was in a decelerating
regime for a long time~\cite{Melchiorri}. In our model,
decelerating and free expanding universe can be obtained from
equation (\ref{FRW2}) for $w_{_{X}}\geq-(1+u)/3$ (where
$u\equiv\rho_{_{M}}/\rho_{_{X}}$) and $0<\alpha\leq1$ in the
power--law solutions. The acceptable domains of $w_{_{X}}$,
$\alpha$ and $\lambda$, in accord with this situation, are given
in Table~$1$ and drown in Fig.~$1$.

\item \textbf{II. Acceleration}\\

Recent observations illustrate that the universe is in an
accelerating phase at the present
epoch~\cite{Perlmutter}--\cite{Carroll}. An accelerating universe
from equation (\ref{FRW2}) makes $w_{_{X}}<-(1+u)/3$ and
$\alpha>1$ for the power--law solutions. The acceptable values of
$w_{_{X}}$, $\alpha$ and $\lambda$ corresponding to this condition
are  also given in Table~$1$ and Fig.~$1$.
\end{description}

Table~$1$ and Fig.~$1$ surprisingly show that both the
decelerating and the accelerating solutions are acceptable in the
range $w_{_{X}}<-1/3$. However, a fixed value for the parameter
$\alpha$ for all epochs implies that the universe is always either
decelerating or accelerating~\cite{Sen}, which is the common
property of the power--law solutions. In this respect, one should
consider the model only for the late time acceleration of the
universe.
\begin{table}
\begin{center}
\begin{tabular}{|c|c|c|c|c|}
  \hline
 $\rm Universe \ Phase$ & $\alpha$ & $\lambda \ \rm for \ case \ A$ & $\lambda \ \rm for \ case \  B$ & $w_{_{X}}$ \\ \hline\hline
$\rm Expantion$ & $0<\alpha\leq2/3$ & $-2<\lambda\leq0$ & $\lambda\leq0$ & $w_{_{X}}\geq(2-3\alpha)/3\alpha$ \\
& $\alpha>2/3$ & $\lambda>0$ & $\lambda>0$ & $w_{_{X}}<(2-3\alpha)/3\alpha$ \\ \hline
$\rm Deceleration \ and$ & $0\leq\alpha\leq2/3$ & $-2<\lambda\leq0$ & $\lambda\leq0$ & $w_{_{X}}\geq(2-3\alpha)/3\alpha$ \\
$\rm Free \ Expansion$ & $2/3<\alpha\leq1$ & $0<\lambda\leq1$ & $\lambda\geq0$ & $w_{_{X}}\leq(2-3\alpha)/3\alpha$ \\ \hline
$\rm Acceleration$ & $\alpha>1$ & $\lambda>1$ & $\lambda>0$ & $w_{_{X}}\leq(2-3\alpha)/3\alpha$ \\ \hline
\end{tabular}
\end{center}
\caption{\footnotesize The ranges of $w_{_{X}}$, $\alpha$ and $\lambda$ for the
expanding, decelerating and accelerating power--law solutions of interacting cases A and B.}
\end{table}
\begin{figure}
\begin{center}
\epsfig{figure=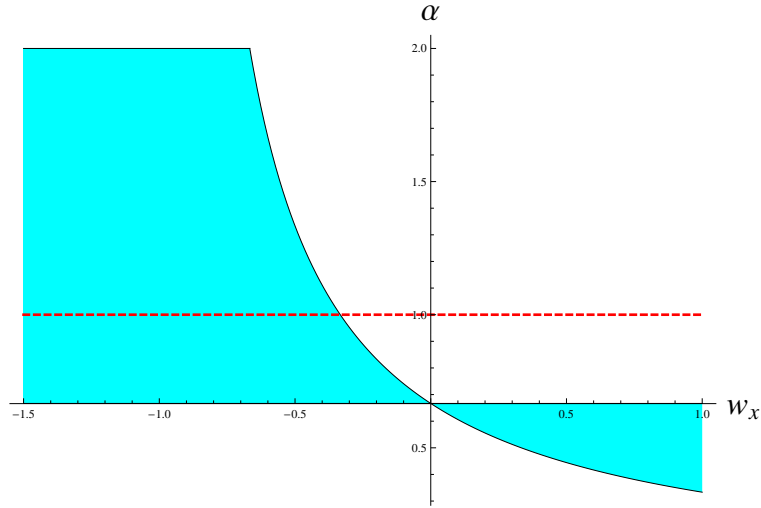,width=10cm} \caption{\footnotesize The
domains of $w_{_{X}}$ and $\alpha$ correspond to Table~$1$. The
dashed line corresponds to the border value $\alpha=1$, that
separates the acceleration (the upper) and the deceleration (the
lower) regions.}
\end{center}
\end{figure}

To analyze the nature of the acceleration part of the model, it is
instructive to investigate the model consistency with the
observational data. For this purpose, in the next section, we
employ the MCMC method to fit the free parameters of the model
with the SCP Union2.1 SN Ia
compilation~\cite{suzuki2011,supernova} based on the Bayesian
statistics~\cite{AmirHajian}.

\section{Observational Constraints}
\indent

The various data information from different observations are used
to constrain the cosmological models. Among them, the SN Ia
distance modulus declares the accelerated expansion of the
universe~\cite{Perlmutter}--\cite{Carroll}. We employ the Bayesian
statistics to investigate the model consistency with the SCP
Union2.1 SN Ia compilation that is an update of the Union2
compilation~\cite{amanullah2010}. This compilation consists of
nineteen data sets from 833 supernovae. However, it has been
indicated~\cite{suzuki2011,supernova} that only 580 of these 833
supernovae can pass the usability cuts, which contain new data
from the HST Cluster Survey. The HST Cluster Survey supplies the
latest and the most complete data set for SN Ia observations till
now.

The most popular techniques for the parameter estimation in
cosmology is the MCMC method. These methods were first employed in
astrophysics~\cite{Christensen}, and since then, it has also been
used in cosmology. We have used our own package, however, the
standard packages for the cosmological parameter estimation are
also publicly available~\cite{Lewis,Doran}.

The likelihood function in the Bayesian statistics is defined to
be proportional to $ exp(-\chi^{2}/2)$. Hence, the best fitted
values for the parameters of the model are obtained by minimizing
the $\chi^{2}$, which for the SN Ia is defined as
\begin{equation}\label{Kai}
\chi^{2}_{_{\rm{SN}}}=\sum_{i=1}^{580}\frac{\Big[\mu_{\rm{th}}(z_{i};\{l\})-\mu_{\rm{ob}}(z_{i})\Big]^{2}}{\sigma_{i}^{2}}\, .
\end{equation}
In equation (\ref{Kai}), $\{l\}$ refers to the parameters of the
model which can be estimated by the data analysis process and
$\sigma_{i}$ stands for the $1\sigma$ uncertainty associated to
the \emph{i}th data point. The symbol $\mu_{\rm{ob}}$ is the
observed distance modulus and $\mu_{\rm{th}}$ is the theoretical
distance modulus that is defined as
\begin{equation}\label{miu}
\mu_{\rm{th}}(z;\{l\})\equiv m-M=5\log D_{L}(z;\{l\})+5\log \left(\frac{c/H_{\circ}}{1 \rm{Mpc}}\right)+25\, ,
\end{equation}
where $m$ is the apparent magnitude, $M$ is the absolute magnitude and
\begin{equation}\label{DL}
D_{L}(z;\{l\})\equiv (1+z)\int_{0}^{z}\frac{H_{\circ} dz'}{H(z';\{l\})}
\, ,
\end{equation}
is the luminosity distance.

We constrain the parameters of the model with the observational
data from the supernovae type Ia, for both interacting cases A and
B of the previous section. The analysis reveals that for the
interacting case B, the parameters $\alpha$ and $\lambda$ are
strictly related and any change for each of them in the MCMC
process yields different non--compatible results. By increasing
the steps of the MCMC process, almost the whole parameter space
will be covered. In other words, there are no preferred values for
the parameters of the model in case B regarding to the SNe Ia
observations, hence, we ignore this case.

The result of the data analysis corresponding to the interacting
case A is as follow. The best fitted values for $\alpha$ and $H_o$
are listed in Table~2, that suggest $\alpha\simeq1.61$ and
$H_{\circ}\simeq 69.4 \ \rm (km/s)/Mpc$ within their own
confidence intervals. Fig.~2 illustrates the likelihood functions
for them. The confidence intervals are also represented in Fig.~3
down. Also, the best fitted value for the deceleration parameter,
$q=-\ddot{a}a/\dot{a}^{2}$, within its confidence intervals is
given in Table~2. The Hubble diagram (the distance modulus in
terms of the redshift) simulated based on this case shows
agreement between the model prediction and the observed data,
Fig.~3 up. The Hubble time corresponding to the best fitted value
of $H_{\circ}$ is $t_{H_{\circ}}\simeq 14.1\times 10^{9} yr$.

Now, we can find the other parameters of the model with the best
fitted value of $\alpha$, namely $\alpha=1.61$. First of all, from
relation (\ref{Lambda1}), one gets $\lambda=2.83$, thus, there are
two choices, either $\lambda=\beta$ or $\lambda=\gamma$ (where for
the latter choice, there are two solutions). Substituting $\alpha$
and $\beta$ (or $\gamma$) into equations (\ref{FRW2}) and
(\ref{FRW3}) gives the best values of $\omega$ and $\gamma$ (or
$\beta$) and then, from relation (\ref{W X}), one gets the best
value for $w_{_{X}}$. The given value of $w_{_{X}}$ from this
procedure must fulfil the acceptable ranges of $w_{_{X}}$
according to Table~1, which for $\alpha=1.61$, it reveals that one
must have $w_{_{X}}\leq-0.58$. Table~3 shows all the three groups
of values for the parameters of the model. The results illustrate
that the consistent values of the parameters of the model are
those listed in the last column, which corresponds to
$\lambda\equiv\gamma$ case. This imposes that the best interacting
term consists of the energy density of the matter multiplied by
the ratio of the time derivative of the scalar field to itself.

The consistent values of the parameters indicate that the scale
factor of the fifth dimension shrinks with the time and the model
has a ghost scalar field with $\omega\simeq-7.75$. Note that, as
described in the Introduction, this mismatching value of $\omega$
with the values obtained from the solar system
observations~\cite{Bertotti2003,fujiBook} is
expected~\cite{banerjee2001}--\cite{Danieletal}. Indeed, this
value of the BD coupling constant actually corresponds to an
imaginary conformal scalar field in the Einstein frame, where the
issue of which conformal frame, the Jordan or the Einstein one, is
physical is a matter of debate~\cite{FaraoniA,FaraoniB}. In
addition, in this model, the obtained dark energy state parameter,
$w_{_{X}}$, is also $\simeq-0.67>-1$ which does~not belong to a
ghost one and the corresponding dark energy does~not lead to big
rip singularities.
\begin{table}
\begin{center}
\begin{tabular}{|c|c|c|c|c|}
  \hline
 $\rm Parameter$ & $\rm Best \ fit$ & $1\sigma$ & $2\sigma$ & $3\sigma$ \\ \hline\hline
$\alpha$ & $1.61$ & $+0.16$ & $+0.29$ & $+0.38$ \\
$$ & $$ & $-0.12$ & $-0.20$ & $-0.25$ \\ \hline
$q_{\circ}$ & $-0.37$ & $\pm0.05$ & $\pm0.09$ & $\pm0.12$ \\ \hline
$H_{\circ} \rm {(km/s)/Mpc}$ & $69.4$ & $+0.5$ & $+0.7$ & $+0.13$ \\
$$ & $$ & $-0.6$ & $-0.9$ & $-0.12$ \\ \hline
\end{tabular}
\end{center}
\caption{\footnotesize The joint likelihood analysis results of
$\alpha$ (with the prior used from $1.00$ to infinity),
$q_{\circ}$ and $H_{\circ}$ up to $3\sigma$ confidence levels for
the interacting case A.}
\end{table}
\begin{figure}
\begin{center}
\epsfig{figure=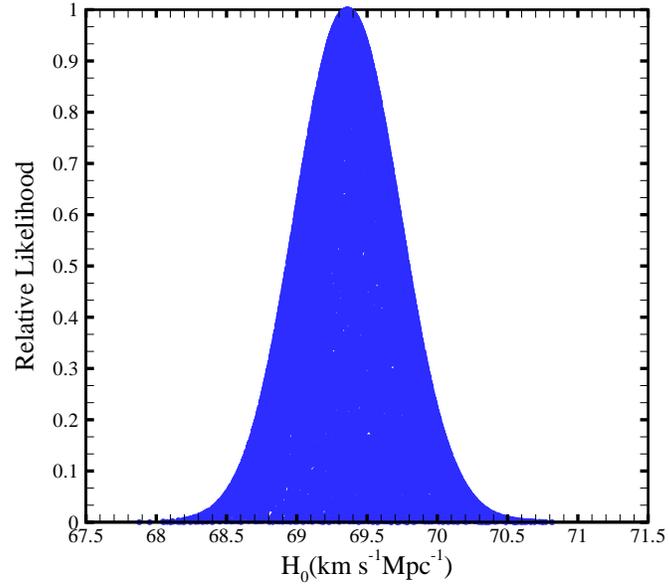,width=10cm}\vspace {4mm}
\epsfig{figure=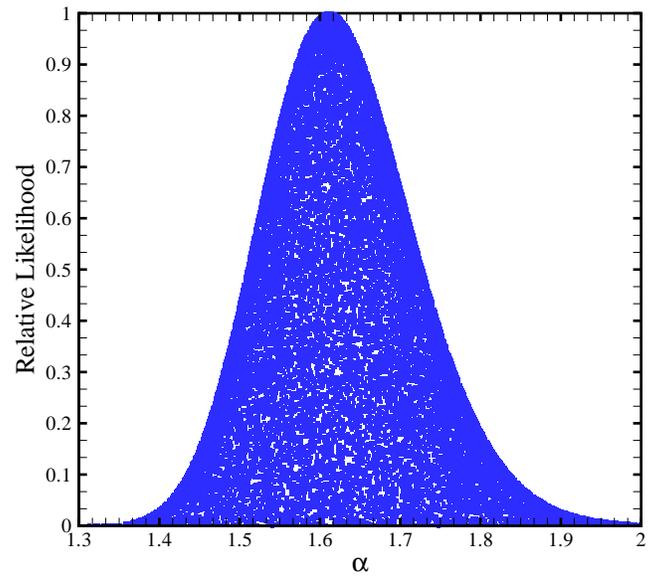,width=10cm} \caption{\footnotesize The
likelihood functions of the model free parameters $H_{\circ}$ and
$\alpha$ for the interacting case A.}
\end{center}
\end{figure}
\begin{figure}
\hspace {3.35cm} \epsfig{figure=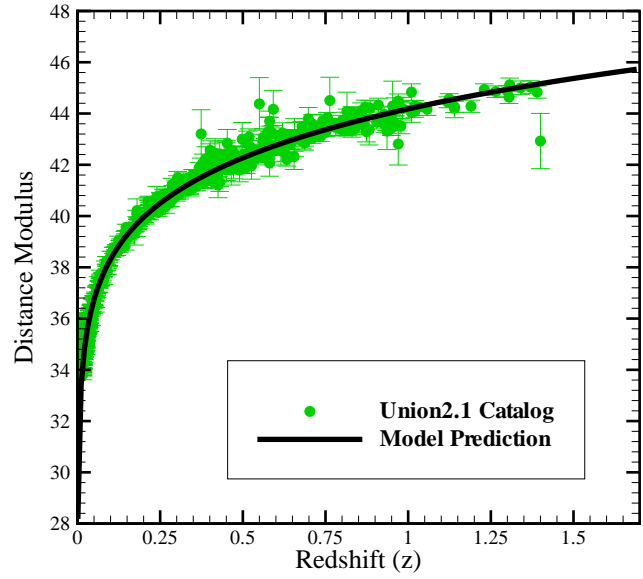,width=10cm}\vspace {4mm}
\begin{center}
\epsfig{figure=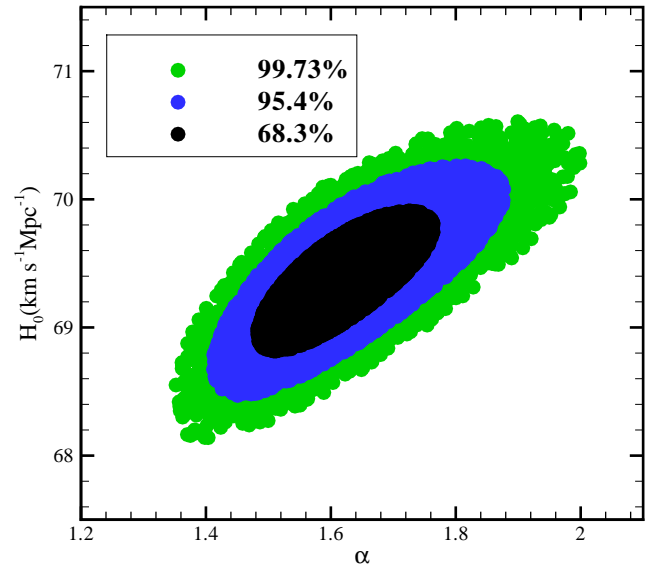,width=10cm} \caption{\footnotesize Upper
figure: the Hubble diagram with simulation based for the
interacting case A and the SCP Union2.1 compilation of the SNe Ia
observations. Lower figure: the joint likelihood analysis of
$\alpha$ and $H_0$ for the interacting case A up to $1\sigma$,
$2\sigma$ and $3\sigma$ levels.}
\end{center}
\end{figure}
\begin{table}
\begin{center}
\begin{tabular}{|c|c|c|c|}
  \hline
$\rm Parameter$ & $\lambda\equiv\beta$ & $\lambda\equiv\gamma$ & $\lambda\equiv\gamma$ \\
$$ & $$ & $\rm Solution \ I$ & $\rm Solution \ II$ \\ \hline\hline
$\beta$ & $2.83$ & $1.74$ & $-6.78$ \\ \hline
$\gamma$ & $-5.74$ & $2.83$ & $2.83$ \\ \hline
$\omega$ & $-1.69$ & $-1.72$ & $-7.75$ \\ \hline
$w_{_{X}}$ & $1.52$ & $0.45$ & $-0.67$ \\ \hline
\end{tabular}
\end{center}
\caption{\footnotesize The best values of the parameters of the
model corresponding to the best fitted value $\alpha=1.61$ for the
interacting case A. The consistent values of the parameters with
the condition $w_{_{X}}\leq(2-3\alpha)/3\alpha=-0.58$ are those
given in the last column.}
\end{table}

\section{Conclusions}
\indent

It is a general belief that the universe is in an accelerated
expanding phase. Thus, the main content of it should be consisted
of what usually has been called dark energy. Hence, a considerable
amount of work has been performed to explain the acceleration of
the universe, but until now, the origin and the nature of dark
energy is unknown. Most of the dark energy models, such as the
quintessence, the chaplygin gas and the k-essence models, involve
minimally coupled scalar fields with different potentials which
have been added by hand, and nothing has been asserted about their
origins. Also in the recent decade, explaining the accelerated
expansion of the universe by alternative theories of gravitation
has been a great challenge.

In this work, following the approach of the induced--matter
theory, we have investigated the cosmological implications of a
$5D$ BD theory, in order to explain the acceleration of the
universe. After inducing in a $4D$ hypersurface, we have
classified the energy--momentum tensor into two parts. One part
represents all kind of the matter (the baryonic and dark), and the
other one contains every extra terms emerging from the scale
factor of the fifth dimension and the scalar field, which we have
considered as the energy--momentum tensor of dark energy.

As the cosmological equations of the model are extremely
non--linearly coupled, we have assumed some simplifications in
order to proceed the properties of the model. The total energy
conservation has been separated into two equations, one for the
matter conservation and the other for dark energy one. Of course,
such a procedure has been performed without interacting term and
with two particular interacting terms between the two parts. We
consider the parameter of the state equation of dark energy to be
constant. Hence, the equations of the model admit the power--law
solutions which impose a spatially flat geometry.

The non--interacting case gives decelerated universes, though, the
interacting cases give both decelerated and accelerated universes.
For the latter case, we have figured out analytically the
acceptable ranges of parameters of the model, and have illustrated
the results in a few tables and figures.

Then after, for these interacting cases, we have employed the MCMC
method based on the Bayesian statistics to investigate the
consistency of the model parameters with the observational data
from supernovae type Ia. For this purpose, we have used the data
of the SCP Union2.1 SN Ia compilation. The data analysis process
for the case, which its interacting term between the matter and
dark energy is proportional to the energy density of dark energy,
reveals that the involved parameters are strictly related and any
change for each of them in the MCMC process yields different
non--compatible results. By increasing the steps of the MCMC
process, almost the whole parameter space will be covered. In
other words, there are no preferred values for the parameters of
the model in this interacting case. But for the other interacting
case, which the interacting term is proportional to the energy
density of the matter, the best fitted values suggest that
$H_{\circ}\simeq 69.4$ (km/s)/Mpc, $q_{\circ}\simeq-0.38$ within
their confidence intervals. The Hubble time corresponding to the
best fitted value of $H_{\circ}$ is $t_{H_{\circ}}\simeq
14.1\times 10^{9} yr$. The best fitted values of the parameters of
the model are also listed in a table, and the results suggest that
the energy density of the matter multiplied by the ratio of the
time derivative of the scalar field to itself plays the role of
the best interacting term between the matter and dark energy.
Also, the consistent values of the parameters indicate that the
model has a ghost scalar field with $\omega\simeq-7.75$, while the
scale factor of the fifth dimension shrinks with the time.
Although the model has a ghost scalar field with this value for
the BD coupling constant, the consistent values of the parameters
reveals a non--ghost dark energy with the state parameter
$w_{_{X}}\simeq-0.67$ as well. That is, with this value of the
dark energy state parameter, there should~not be big rip
singularities and ghost instabilities.

\section*{Acknowledgements}
\indent

We would like to thank Dr. M.S. Movahed for useful comments. M.F.
also thanks the Research Office of Shahid Beheshti University G.C.
for financial support.

%
\end{document}